          [2001/04/25 1.1 (PWD)]
\documentclass{aa}

\usepackage{graphicx,color}
\usepackage{txfonts}
\usepackage{natbib}
\bibpunct[, ]{(}{)}{;}{a}{}{,}
\begin{document}

\title{The habitability  of super-Earths in Gliese~581}
\author{W. von Bloh
\inst{1}
\and
C. Bounama\inst{1}
\and
M. Cuntz\inst{2}
\and
S. Franck\inst{1}
}
\offprints{W. von Bloh}
\institute{Potsdam Institute for Climate Impact Research, P.O. Box 60 12 03, 14412 Potsdam, Germany\\
\email{bloh@pik-potsdam.de}
\and
Department of Physics, University of Texas at Arlington, Box 19059, Arlington, TX 76019, USA}
\date{Received  / Accepted }
\abstract{}
{The planetary system around the M star Gliese 581 consists of a hot Neptune (Gl 581b) and two 
super-Earths (Gl 581c and Gl 581d).  The habitability of this system with respect to the super-Earths 
is investigated following a concept that studies the long-term possibility of photosynthetic
biomass production on a dynamically active planet.}
{A thermal evolution model for a super-Earth is used to calculate the
sources and sinks of atmospheric carbon dioxide.  The habitable zone is determined
by the limits of photosynthetic life on the planetary surface.  Models with different ratios
of land / ocean coverage are investigated.}
{The super-Earth Gl 581c is clearly outside the habitable zone, since it is too close to the
star.  In contrast, Gl 581d is a tidally locked habitable super-Earth near the outer edge
of the habitable zone.  Despite the adverse conditions on this planet, at least some primitive
forms of life may be able to exist on its surface. Therefore, Gl 581d is an interesting target
for the planned TPF/Darwin missions to search for biomarkers in planetary atmospheres.}
{}

\keywords{stars: individual: Gl 581 --- stars: planetary systems --- astrobiology}

\maketitle

%%%%%%%%%%%%%%%%%%%%%%%%%%%%%%%%%%%%%%%%%%%%%%%%%%%%%%%%%%%%%%%%%%%%%%%%%%%%%%%%%%%%%%%%%%%

\section{Introduction}

Planets have now been observed around more than 200 main-sequence stars.  Based on
the available observational techniques, most detected objects are giant (Jupiter-like)
planets.  Therefore, until very recently the existence and possible habitability of Earth-like
planets in extrasolar planetary systems was highly speculative \citep{cuntz03,franck03,
vonbloh03a,jones06,vonbloh07}.
\cite{bonfils05} reported the detection of a Neptune size planet around Gl~581, an M dwarf star at a distance of 6.26 pc
with a mass of $0.31 M_{\odot}$ and a luminosity of 0.013~$L_{\odot}$.  Very recently, \cite{udry07}
announced the detection of two so called ``super-Earth" planets in this system, Gl~581c with a mass of
5.06~$M_\oplus$ with a semi-major axis of 0.073 AU, and Gl~581d with 8.3~$M_\oplus$ and
0.25 AU.  Both mass estimates are minimum masses uncorrected for the inclination term $\sin i$,
which is currently unknown.

According to \cite{valencia06}, super-Earths are rocky planets from one to ten Earth masses with
the same chemical and mineral composition as the Earth.  In the following, we adopt the hypothesis
that this is indeed the case, and consider for these planets our model previously developed for
the Earth, using when appropriate scaling laws.  This justifies the term super-Earth used here
for these planets.

The main question is whether any of the two super-Earths around Gl 581 can harbour life,
i.e., that any of the planets lies within the habitable zone (HZ).  
 As a first approximation, \cite{udry07} computed an equilibrium surface temperature for
Gl 581c of $20^{\circ}$C for an albedo of $0.5$.  They neglected, however, the likely
greenhouse effect of the atmosphere. 
Typically, stellar HZs
are defined as regions around the central star, where the physical conditions are favourable
for liquid water to exist at the planet's surface for a period of time long enough for
biological evolution to occur.  \cite{kasting93} calculated the HZ boundaries for the
luminosity and effective temperature of the present Sun as $R_{\mathrm{in}} = 0.82$ AU
and $R_{\mathrm{out}} = 1.62$ AU.  They defined the HZ of an Earth-like planet as the region
where liquid water is present at the surface.

According to this definition, the inner boundary
of the HZ is determined by the loss of water via photolysis and hydrogen escape.  The outer
boundary of the HZ is determined by the condensation of CO$_2$ crystals out of the atmosphere
that attenuate the incident sunlight by Rayleigh scattering.  The critical CO$_2$ partial
pressure for the onset of this effect is about 5 to 6 bar.  However,
the cooling effect of CO$_2$ clouds has been challenged by \cite{forget97}.
CO$_2$ clouds have the additional effect of reflecting the outgoing thermal
radiation back to the surface.  The precise inner and outer limits of the
climatic habitable zone are still unknown due to the limitations of the
existing climate models.  For the present Sun, the HZ is probably smaller than
between 0.7 to 2 AU, but it is still impossible to give a better constraint,
particularly for the outer boundary of the HZ.  For limitations of the
planetary habitability of M-type stars see \cite{tarter07}.

The luminosity and age of the central star play important roles in the
manifestation of habitability.  The luminosity of Gl 581 can be obtained by
(1) photometry \citep{bonfils05,udry07}, and (2) the application of the
mass-radius relationship \citep{ribas06} together with the spectroscopically
determined stellar effective temperature of $T_e = 3480$~K
\citep{bean06}.  Both methods yield $L=0.013 \pm 0.002 L_\odot$.
\cite{bonfils05} consider a stellar age of at least 2 Gyr.

In the following, we adopt a definition of the HZ previously used by
\cite{franck00a,franck00b}.  Here habitability
at all times does not just depend on the parameters of the central star, but
also on the properties of the planet.  In particular, habitability is linked
to the photosynthetic activity of the planet, which in turn depends on the
planetary atmospheric CO$_2$ concentration together with the presence of liquid water, and is thus strongly influenced by
the planetary dynamics.  
We call this definition the photosynthesis-sustaining habitable zone, pHZ.  
In principle, this leads to additional spatial {\it and
temporal} limitations of habitability, as the pHZ (defined for a specific
type of planet) becomes narrower with time due to the persistent decrease of
the planetary atmospheric CO$_2$ concentration. 

%%%%%%%%%%%%%%%%%%%%%%%%%%%%%%%%%%%%%%%%%%%%%%%%%%%%%%%%%%%%%%%%%%%%%%%%%%%%%%%%%%%%%%%%%%%

\section{Estimating the habitability of a super-Earth}

\subsection{Definition of the photosynthesis-sustaining habitable zone}

The climatic habitable zone at a given time for a star with luminosity $L$ and
effective temperature $T_e$ different from the Sun can be calculated
according to \cite{jones06} based on previous results by \cite{kasting93} as
\begin{equation}
R_{\mathrm{in}} = \left(\frac{L}{L_\odot\cdot S_{\mathrm{in}}(T_e)}\right)^{\frac{1}{2}}~,~~R_{\mathrm{out}}
                = \left(\frac{L}{L_\odot\cdot S_{\mathrm{out}}(T_e)}\right)^{\frac{1}{2}}
\label{hz_jones}
\end{equation}
with $S_{\mathrm{in}}(T_e)$, $S_{\mathrm{out}}(T_e)$ given as second order polynomials.

To assess the habitability of a terrestrial planet, an Earth-system model
is applied to calculate the evolution of the temperature and 
atmospheric CO$_2$ concentration.  On Earth, the carbonate-silicate cycle
is the crucial element for a long-term homeostasis under increasing solar
luminosity.  On geological time-scales, the deeper parts of the Earth are
considerable sinks and sources of carbon.  
\begin{figure}
\centering
\resizebox{0.7 \hsize}{!}{\includegraphics[width=1cm]{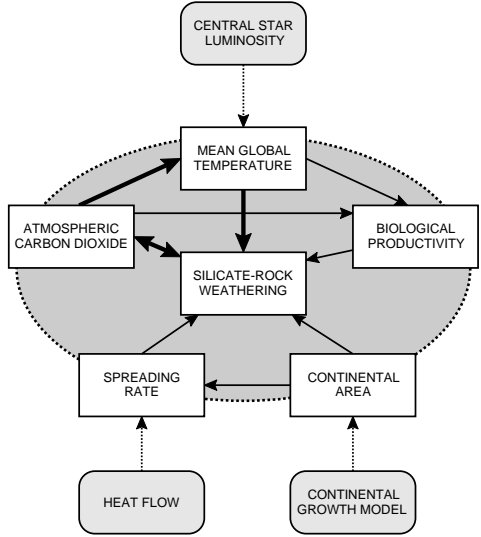}}
\caption{Earth system box model.  The arrows indicate the different forcing
and feedback mechanisms.  The bold arrows indicate negative feedback operating
towards climate stabilization.
} \label{block}
\end{figure}

Our numerical model couples the stellar luminosity $L$, the silicate-rock
weathering rate $F_{\mathrm{wr}}$ and the global energy balance to obtain
estimates of the partial pressure of atmospheric carbon dioxide
$P_{\mathrm{CO}_2}$, the mean global surface
temperature $T_{\mathrm{surf}}$, and the biological productivity $\Pi$ as
a function of time $t$ (Fig.~\ref{block}).  The main point is the
persistent balance between the CO$_2$ sink in the atmosphere-ocean system and
the metamorphic (plate-tectonic) sources.  This is expressed through the
dimensionless quantities  
\begin{equation}
f_{\mathrm{wr}}(t) \cdot f_A(t) = f_{\mathrm{sr}}(t),
\label{gfr}
\end{equation}
where $f_{\mathrm{wr}}(t) \equiv F_{\mathrm{wr}}(t)/F_{\mathrm{wr},0}$ is the 
weathering rate, $f_A(t) \equiv A_c(t)/A_{c,0}$ is the continental area, and
$f_{\mathrm{sr}}(t) \equiv S(t)/S_0$ is the areal spreading rate, which are all
normalized by their present values of Earth.  
Eq.~(\ref{gfr}) can be rearranged by introducing the geophysical forcing ratio,
GFR \citep{volk87} as
\begin{equation}
f_{\mathrm{wr}}(T_{\mathrm{surf}},P_{\mathrm{CO}_2})=\frac{f_{\mathrm{sr}}}{f_A} \ =: \ \mathrm{GFR}(t)
.
\label{gfr2}
\end{equation}
Here we assume that the weathering rate depends only on the global surface temperature
and the atmospheric CO$_2$ concentration.
For the investigation of a super-Earth under external forcing,
we adopt a model planet with a prescribed continental area.  The fraction of
continental area relative to the total planetary surface $f_A$ is varied between
$0.1$ and $0.9$.

The connection between the stellar parameters and the planetary climate can be
formulated by using a radiation balance equation \citep{williams98}
\begin{equation}
\frac{L}{4\pi R^2} [1- a (T_{\mathrm{surf}}, P_{\mathrm{CO}_2})]
 = 4I_R (T_{\mathrm{surf}}, P_{\mathrm{CO}_2}),
\label{L}
\end{equation}
where $a$ denotes the planetary albedo, $I_R$ the outgoing infrared flux, and $R$ the distance from the central star. 
The Eqs.~(\ref{gfr2}) and (\ref{L}) constitute a set of two coupled equations with two unknowns,
$T_{\mathrm{surf}}$ and $P_{\mathrm{CO}_2}$, if the parameterisation of the weathering rate, the luminosity, the
distance to central star and the geophysical forcing ratio are specified.  Therefore, a numerical solution can be
attained in a straightforward manner.

The photosynthesis-sustaining HZ around Gl~581 is defined as the spatial domain of all distances $R$ from
the central star where the biological productivity is greater than zero, i.e.,
\begin{equation}
{\mathrm{pHZ}} \ := \ \{ R \mid \Pi (P_{\mathrm{CO}_2}(R,t), T_{\mathrm{surf}}(R,t))>0 \}.
\label{hz}
\end{equation}
In our model, biological productivity is considered to be solely a function of
the surface temperature and the CO$_2$ partial pressure in the atmosphere.
Our parameterisation yields maximum productivity at $T_{\mathrm{surf}} = 50^{\circ}$C
and zero productivity for $T_{\mathrm{surf}} \leq 0^{\circ}$C or $T_{\mathrm{surf}}
\geq 100^{\circ}$C or $P_{\mathrm{CO}_2}\leq 10^{-5}$ bar \citep{franck00a}.
The inner and outer boundaries of the pHZ do not depend on
the detailed parameterisation of the biological productivity within the temperature
and pressure tolerance window.  Hyperthermophilic life forms can tolerate
temperatures somewhat above $100^{\circ}$C.  However, these chemoautotrophic organisms are outside the
scope of this study.

\subsection{Silicate rock weathering}

Weathering plays an important role in Earth's climate because it provides the main sink for
atmospheric carbon dioxide. The overall chemical reactions for the weathering process are
\begin{eqnarray*}
\mbox{CO$_2$}+ \mbox{CaSiO$_3$} &\rightarrow & \mbox{CaCO$_3$} + \mbox{SiO$_2$} \\
\mbox{CO$_2$}+ \mbox{MgSiO$_3$} & \rightarrow & \mbox{MgCO$_3$} + \mbox{SiO$_2$}
\end{eqnarray*}
The total process of weathering embraces (1) the reaction of silicate minerals with carbon dioxide,
(2) the transport of weathering products, and (3) the deposition of carbonate minerals in sediments.
When combining all these effects, the normalized global mean weathering rate $f_{\mathrm{wr}}$ can be
calculated as
\begin{equation}
f_{\mathrm{wr}} = {\left( \frac{a_{\mathrm{H}^+}}{a_{\mathrm{H}^+,0}}\right)}^{0.5}
     \exp \left( \frac{T_{\mathrm{surf}}-T_{\mathrm{surf},0}}{13.7 \mathrm{K}} \right).\label{hz:eq1}
\end{equation}
following \cite{walker81}.
Here the first factor reflects the role of the CO$_2$ concentration in the soil, $P_{\mathrm{soil}}$,
with $a_{\mathrm{H}^+}$ as the activity of $\mathrm{H}^+$ in fresh soil-water that depends on
$P_{\mathrm{soil}}$ and the global mean surface temperature $T_{\mathrm{surf}}$. The quantities
$a_{\mathrm{H}^+,0}$ and $T_{\mathrm{surf},0}$ are the present-day values for the $\mathrm{H}^+$
activity and the surface temperature, respectively.
The activity $a_{\mathrm{H}^+}$ is itself a function of the temperature and the CO$_2$ concentration of the soil.
The equilibrium constants for the chemical activities of the carbon and sulfur systems involved are
taken from \cite{stumm81}. Note that the sulfur content of the soil also contributes to the
global weathering rate, but its influence does not depend on the temperature. It can be regarded as
an overall weathering attribute that has to be taken into account for the estimation of the
present-day value.

For any given weathering rate,
the surface temperature and the CO$_2$ concentration of the soil can be calculated in a self-consistent
manner. $P_{\mathrm{soil}}$ is assumed to be linearly related to
the terrestrial biological productivity $\Pi$ \citep[see][]{volk87} and the atmospheric CO$_2$ concentration
$P_{\mathrm{CO}_2}$. Thus we have
\begin{equation}
\frac{P_{\mathrm{soil}}}{P_{\mathrm{soil},0}} = \frac{\Pi}{\Pi_0} \left( 1- \frac{P_{\mathrm{CO}_2,0}}{P_{\mathrm{soil},0}} \right)
     + \frac{P_{\mathrm{CO}_2}}{P_{\mathrm{soil},0}} ,\label{hz:eq2}
\end{equation}
where $P_{\mathrm{soil},0}$, $\Pi_0$ and $P_{\mathrm{CO}_2,0}$ are again present-day values.

Since Earth is now harbouring complex life, the weathering rates might be lower on
a planet without a complex biosphere.  The parameterisation of biotic enhancement of weathering is based only on an increase
of the CO$_2$ concentration in the soil.  This implies a rather weak functional dependence of the weathering rate
on biological productivity.  A ten-fold increase in soil CO$_2$ concentration relative to the atmosphere only amounts to a 1.56
fold increase in the weathering rate due to the present biota.  This is a significant underestimate, indicating that
much of the observed biotic amplification of weathering is due to processes other than increased
soil CO$_2$ concentration.  According to \cite{schwartzman99} the total amplification due to complex land life is at least a
factor of 10 and may exceed 100. To explore the effect of a stronger biological amplification of weathering
a direct dependence \citep{lenton01} of weathering rate, $f_{\rm{wr}}$, on
productivity of complex life (e.g. land plants), $\Pi_{\mathrm{complex}}$, with amplification factor, $\alpha_{\mathrm{bio}}$, can be included:
\begin{equation}
f'_{\rm{wr}}=\left(\left(1-\frac{1}{\alpha_{\mathrm{bio}}}\right) \label{alpha}
\frac{\Pi_{\mathrm{complex}}}{\Pi_{\mathrm{complex},0}}+\frac{1}{\alpha_{\mathrm{bio}}}\right)f_{\mathrm{wr}},\label{eq:enhance}
\end{equation}
where $\Pi_{\mathrm{complex,0}}$ is the productivity of the present biosphere.
The aim of this calculation is to obtain the value for weathering for a planet
without a complex biosphere. Therefore the weathering rate for an Earth with primitive life
 $f_{\mathrm{wr,primitive}}$
($\Pi_{\mathrm{complex}}\equiv 0$) is a factor of $\alpha_{\mathrm{bio}}$ lower than
the weathering rate for a complex biosphere
\begin{equation}
f_{\mathrm{wr,primitive}}=\frac{1}{\alpha_{\mathrm{bio}}} f_{\mathrm{wr}}, \alpha_{\mathrm{bio}}>1. \label{eq:dc2}
\end{equation}

\subsection{Thermal evolution model}

Parameterised convection models are the simplest models for investigating the thermal
evolution of terrestrial planets and satellites. They have been successfully applied to the 
evolution of Mercury, Venus, Earth, Mars, and the Moon \citep{stevenson83,sleep00}.
\cite{franck95} have investigated the thermal and volatile history of Earth and Venus in the
framework of comparative planetology. The internal structure of massive terrestrial planets
with one to ten Earth masses has been investigated by \cite{valencia06} to
obtain scaling laws for total radius, mantle thickness, core size, and average density as 
a function of mass. Similar scaling laws were found for different compositions. We will
use such scaling laws for mass-dependent properties of super-Earths and also mass-independent
material properties given by \cite{franck95}.

The thermal history and future of a super-Earth has to be determined to
calculate the spreading rate for solving key Eq.~(\ref{gfr}).
A parameterised model of whole mantle convection including the volatile exchange
between the mantle and surface reservoirs \citep{franck95,franck98} is applied.
Assuming conservation of energy, the average mantle temperature $T_m$ can be
obtained as
\begin{equation} 
{4 \over 3} \pi \rho c (R_m^3-R_c^3) \frac{dT_m}{dt} = -4 \pi
R_m^2 q_m + {4 \over 3} \pi E(t) (R_m^3-R_c^3), \label{therm} \end{equation}
where $\rho$ is the density, $c$ is the specific heat at constant pressure,
$q_m$ is the heat flow from the mantle, $E(t)$ is the energy production rate by
decay of radiogenic heat sources in the mantle per unit volume, and $R_m$ and
$R_c$ are the outer and inner radii of the mantle, respectively. The radiogenic
heat source per unit volume is parameterised as
\begin{equation}
E(t)=E_0e^{-\lambda t}
\end{equation}
where $\lambda$ is the decay constant of radiogenic heat and the constant $E_0$
is obtained from the present heat flux of $q_m=0.07$ Wm$^{-2}$
for an Earth-size planet at 4.6 Gyr.

The mantle heat flow is parameterised in terms of the Rayleigh number $\mathrm{Ra}$ as
\begin{equation}
q_m = {k (T_m - T_{\mathrm{surf}}) \over R_m -R_c} \left({\mathrm{Ra} \over
\mathrm{Ra}_{\rm{crit}}}\right)^\beta, \label{eqheat}
\end{equation}
with
\begin{equation}
\mathrm{Ra} = {g \alpha (T_m - T_{\mathrm{surf}}) (R_m - R_c)^3 \over \kappa \nu},
\label{eqrayleigh}
\end{equation}
where $k$ is the thermal conductivity, $\mathrm{Ra}_{\rm{crit}}$ is the critical value
of $\mathrm{Ra}$ for the onset of convection, $\beta$ is an empirical constant, $g$ is
the gravitational acceleration, $\alpha$ is the coefficient of thermal expansion,
$\kappa$ is the thermal diffusivity, and $\nu$ is the water-dependent
kinematic viscosity. The viscosity $\nu$ can be calculated with the help of a water
fugacity dependent mantle creep rate. It strongly depends on the evolution of 
the mass of mantle water, $M_w$, and the mantle temperature, $T_m$, i.e.,
$\nu\equiv\nu(T_m,M_w)$ and is parameterised according to \cite{franck95}. 

The evolution of the mantle water
can be described by a balance equation between the regassing flux $F_{\mathrm{reg}}$
and outgassing flux $F_{\mathrm{out}}$ as
\begin{eqnarray}
\frac{dM_w}{dt} & = & F_{\mathrm{reg}}-F_{\mathrm{out}}  \nonumber \\
 & = & f_{\mathrm{bas}}\rho_{\mathrm{bas}}d_{\mathrm{bas}}SR_{\mathrm{H_2O}}-\frac{M_w}
         {\frac{4}{3}\pi(R_m^3-R_c^3)}d_mf_wS,
\end{eqnarray}
where
$f_{\mathrm{bas}}$ is the water content in the basalt layer,  
$\rho_{\mathrm{bas}}$ is the average density,
$d_{\mathrm{bas}}$ is the average thickness of the
basalt layer before subduction,
$S$ is the areal spreading rate,
$d_m$ is the melting generation depth and $f_w$ is the outgasssing fraction of water.
$R_{\mathrm{H_2O}}$ is the regassing ratio of water, i.e., the fraction of subducting water
that actually enters the deep mantle.  The regassing ratio depends linearly on the mean
mantle temperature $T_m$ that is derived from the thermal evolution model via
\begin{equation} 
R_{\mathrm{H_2O}}(T_m)=R_T \cdot\left(T_m(0)-T_m\right)+R_{\mathrm{H_2O},0}.\label{eq5} 
\end{equation}
The factor $R_T$ is adjusted to get the correct modern amount of surface water
(one ocean mass) for an Earth-size planet and $R_{\mathrm{H_2O},0}$ is fixed at $0.001$,
i.e., the value is very
low at the beginning of the planetary evolution  because of the enhanced loss of
volatiles resulting from back-arc volcanism at higher temperatures.

The areal spreading rate $S$ is a function of the average mantle temperature $T_m$, the 
surface temperature $T_{\mathrm{surf}}$, the heat flow from the mantle $q_m$, and the
area of ocean basins $A_0$ \citep{turcotte82}, given as
\begin{equation}  S = \frac{q_m^2 \pi
\kappa A_0}{4 k^2 (T_m - T_\mathrm{surf})^2}\,. 
\end{equation}
In order to calculate the spreading rates for a planet with several Earth masses, 
the planetary parameters have to be adjusted.  Therefore, we assume
\begin{equation}
\frac{R_p}{R_{\oplus}}= \left(\frac{M}{M_{\oplus}}\right)^{0.27}
\end{equation} and
where $R_p$ is the planetary radius, see \cite{valencia06}.
The total radius, mantle thickness, core size and average density are all functions
of mass, with subscript $\oplus$ denoting Earth values.
The exponent of $0.27$ has been obtained for super-Earths ($M>1 M_\oplus$).  The values
of $R_m$, $R_c$, $A_0$, the density of the planet, and the other planetary
properties are scaled accordingly.

The CO$_2$ concentration in the atmosphere $P_{\mathrm{CO}_2}$ is derived from the total mass of atmospheric carbon $C_{\mathrm{atm}}$ calculated from the balance between sources and sinks according to 
\begin{equation}
P_{\mathrm{CO}_2}=\frac{g}{4\pi R_p^2}\frac{\mu_{\mathrm{CO}_2}}{\mu_{\mathrm{C}}}C_{\mathrm{atm}},
\end{equation}
where $\mu_{\mathrm{CO}_2}$ and $\mu_{\mathrm{C}}$ are the molar weights of CO$_2$ and C,
respectively. The mass dependent pre-factor $g/R_p^2$ scales with $M^{-0.08}\approx M^0$ and has therefore been neglected in our study.

In Tab.~\ref{param} we give a summary of the selected values for the parameters used in the
thermal evolution model of the $5 M_\oplus$ and $8 M_\oplus$ super-Earth planets.
For comparison, the values for an Earth-size planet are also shown. According to
\cite{valencia07}, we assume that a more massive planet is likely to convect in a plate
tectonic regime similar to Earth. Thus, the more massive the planet is, the higher the
Rayleigh number that controls convection, the thinner the top boundary layer (lithosphere),
and the faster the convective velocities.  In a first order approximation, we assume a
fixed thickness of the basalt layer and melting depth corresponding to relatively low values.
Furthermore, the initial amount of water $M_w(0)$ scales 
linearly with the planetary mass. This might be an underestimate because more massive planets tend to accrete more volatiles.
\begin{table*}
\caption{Parameter values for the evolution model for mantle temperature  and water}             % title of Table
\label{param}      % is used to refer this table in the text
\centering                          % used for centering table

\begin{tabular}{l l l l l l}        % centered columns (4 columns)
\hline\hline                 % inserts double horizontal lines
Parameter & \multicolumn{3}{c}{Value}   & Unit & Description \\    % table heading
 ...      & $1 M_\oplus$ & $5 M_\oplus$ & $8 M_\oplus$ & ...  & ...         \\
\hline                        % inserts single horizontal line
$d_{\mathrm{bas}}$    & $5\times 10^3$ & $5\times 10^3$ & $5\times 10^3$ & m & average thickness of the basalt layer \\
$f_{\mathrm{bas}}$    & $0.03$ & $0.03$ & $0.03$ & ... & mass fraction of water in the basalt layer \\
$f_w$                 & $0.194$ & $0.194$ & $0.194$ & ... & degassing fraction of water \\
$d_m$                 & $40\times 10^3$ & $40\times 10^3$ & $40\times 10^3$ & m & melting depth \\
$k$                   & $4.2$ & $4.2$ & $4.2$ & J s$^{-1}$ m$^{-1}$ K$^{-1}$ & thermal conductivity \\
$M_w(0)$              & $4.2\times 10^{21}$ & $2.1\times 10^{22}$ & $5.36\times 10^{22}$ & kg & initial amount of mantle water \\
$R_c$                 & $3,471\times 10^3$ & $5,360\times 10^3$ & $6,085\times 10^3$ & m & inner radius of the mantle \\
$R_m$                 & $6,271\times 10^3$ & $9,684\times 10^3$ & $10,994\times 10^3$ & m & outer radius of the mantle \\
$T_m(0)$              & $3,000$ & $3,000$ & 3,000 & K & initial mantle temperature \\
$\kappa$              & $10^{-6}$ & $10^{-6}$ & $10^{-6}$ & m$^2$ s$^{-1}$ & thermal diffusivity \\
$\rho_{\mathrm{bas}}$ & $2,950$ & $4,005$ & 4,379 & kg m$^{-3}$ & density of the basalt \\
$\rho c$              & $4.2\times 10^6$ & $4.2\times 10^{6}$ & $4.2\times 10^6$ & J m$^{-3}$ K$^{-1}$ & density $\times$ specific heat \\
$R_T$                 &  $29.8\times 10^{-5}$ & $29.8\times 10^{-5}$ &  $29.8 \times 10^{-5}$ & K$^{-1}$ & temperature dependence of regassing ratio \\
$\alpha$              & $3\times 10^{-5}$ & $3\times 10^{-5}$ &$3\times 10^{-5}$ & K$^{-1}$ & coefficient of thermal expansion \\
$\beta$               & $0.3$ & $0.3$ &$0.3$ & ... & empirical constant in Eq. (\ref{eqheat}) \\
$\mathrm{Ra}_{\mathrm{crit}}$ & $1,100$  &$1,100$ &$1,100$ & ... & critical Rayleigh number \\
$\lambda$             & $0.34$ & $0.34$& $0.34$ & Gyr$^{-1}$ & decay constant of radiogenic heat \\
$E_0$                 &  $1.46\times 10^{-7}$ & $1.46\times 10^{-7}$ & $1.46\times 10^{-7}$ & J s$^{-1}$ m$^{-3}$ & initial heat generation per time and volume \\
$g$                   & 9.81 & 20.6 & 25.5 & m s$^{-2}$ & gravitational acceleration \\
\hline                                   %inserts single line
\end{tabular}

\end{table*}

\begin{figure}[h]
\centering
\resizebox{0.78 \hsize}{!}{\includegraphics{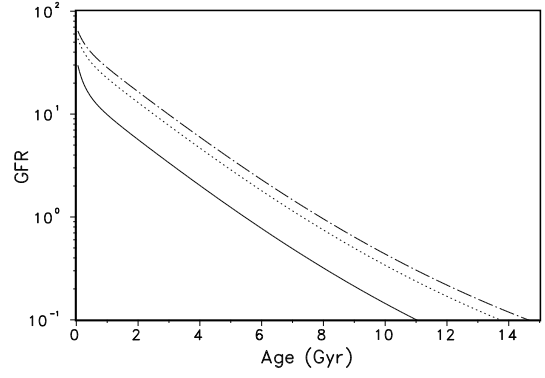}}
\caption{Geophysical forcing ratios (GFR) for a planet with $M=1 M_\oplus$ (solid line), $M=5M_\oplus$ (dotted line) and $M=8M_\oplus$
(dash-dotted line) for a relative continental area of $0.3$.}
\label{fig_gfr}
\end{figure}

The geophysical forcing ratios calculated from the thermal evolution model for a planet with one, five and eight Earth masses
for a relative continental area of $0.3$ are depicted in Fig.~\ref{fig_gfr}. It is obvious that the GFR increases with planetary mass
and is at time zero about 1.8 times higher for $5 M_\oplus$ and 2.1 times higher for $8 M_\oplus$ than for the Earth itself.

\subsection{Tidal locking}

According to \cite{peale77}, the tidal locking radius $r_T$ for a planet on a
circular orbit can be estimated via
\begin{equation}
r_T=0.027 \left( \frac{P_0t}{Q}\right)^{\frac{1}{6}}M_{\mathrm{star}}^{\frac{1}{3}} , \label{tidal}
\end{equation}
where $P_0$ is the original rotation period of the planet, $t$ is the time, $Q^{-1}$ is
the dissipation function and $M_{\mathrm{star}}$ is the stellar mass (all quantities in
cgs units).  We assume analogously to \cite{kasting93} $Q=100$ and $P_0=13.5$ hr.

Planets inside the habitable zone of M stars are tidally locked. 
Due to tidal locking, a weaker intrinsic magnetic field is expected.
Using simple scaling laws for the planetary magnetic dipole moment $m$, summarised by
\cite{griessmeier05}, we estimate values of about 0.5 $m_\oplus$ and 0.1 $m_\oplus$
for Gl 581c and Gl 581d, respectively.  The corresponding sizes of expected magnetospheres
(compressed by coronal winds) can be quantified by the standoff distance 
$R_s$ of the magnetopause \citep{khodachenko07}.  The corresponding values for $R_s$
are on the order of several planetary radii.  Thereby, the surfaces of the super-Earth
planets Gl 581c and Gl 581d are expected to be protected from hot coronal winds. 

For these estimates we assume that the super-Earths Gl 581c and Gl 581d have at least
liquid outer cores.  In contrast, according to \cite{valencia06}, all super-Earth planets
with masses higher than 1 $M_\oplus$ have completely solid cores for the preferred set
of thermodynamic data.  Nevertheless, using so-called high thermal parameters give warmer
interiors and allow super-Earth planets to have a completely liquid core.  This again
points to a magnetospheric protection of super-Earth planetary atmospheres in Gl 581
from a dense flow of stellar plasma. 

Furthermore, low mass stars have strong XUV irradiations during long time periods.
Inside the habitable zone, such radiations can erode the atmosphere of an Earth-size planet
and result in an additional limitation on the definition of habitability \citep{griessmeier05}. 
Another problem for the habitability of tidally locked planets is the freezing out of atmospheric
volatiles on the dark side of the planet that makes the planets not habitable.  Detailed
investigations with the help of three-dimensional global circulation climate models, including
the hydrological cycle, by \cite{joshi97} and \cite{joshi03} showed that approximately 100 mbars
of CO$_2$ are sufficient to prevent atmospheric collapse.  Therefore, super-Earth planets that
generally are assumed to have more volatiles should not be restricted in their habitability
by such effects. In case of an eccentric orbit of Gl 581d ($e=0.2$), the planet might be trapped in a
$m:n$ tidal locking (similar to Mercury in the Solar System), thus increasing its rotational period.

%%%%%%%%%%%%%%%%%%%%%%%%%%%%%%%%%%%%%%%%%%%%%%%%%%%%%%%%%%%%%%%%%%%%%%%%%%%%%%%%%%%%%%%%%%%

\section{Results and discussion}

\begin{figure}[h]
\centering
\resizebox{0.87 \hsize}{!}{\includegraphics{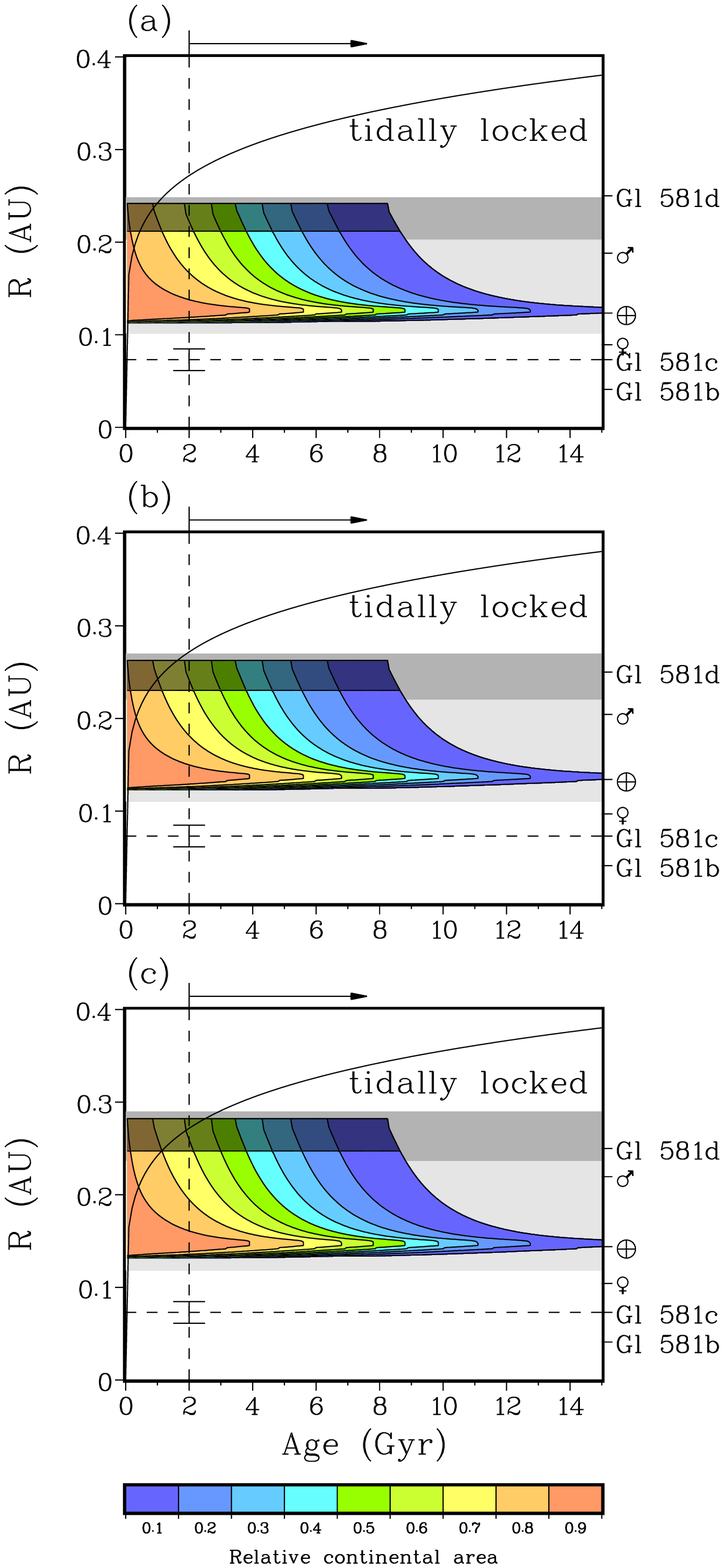}}
\caption{The pHZ of Gl 581 for a super-Earth ($M=5 M_\oplus$) with a relative
continental area varied from 0.1 to 0.9 and a fixed stellar luminosity of (a) $0.011L_\odot$,
(b) $0.013 L_\odot$, and (c) $0.015 L_\odot$ as a function of planetary age. The light colours
correspond to a maximum
CO$_2$ pressure of 5 bar, whereas the dark colours correspond to 10 bar.  For comparison,
the positions of Venus, Earth and Mars are shown scaled to the luminosity of Gl~581.
The light grey shaded area denotes the HZ calculated from Eq.~(\ref{hz_jones}), while the
dark shaded area corresponds to an extended outer limit following \cite{mischna00}.
The vertical bar at 2 Gyr denotes the range of distances due to the (possibly) eccentric
orbit.  The area below the solid black curve is affected by tidal locking.}
\label{fig2}
\end{figure}
\begin{figure}[h]
\centering
\resizebox{0.87 \hsize}{!}{\includegraphics{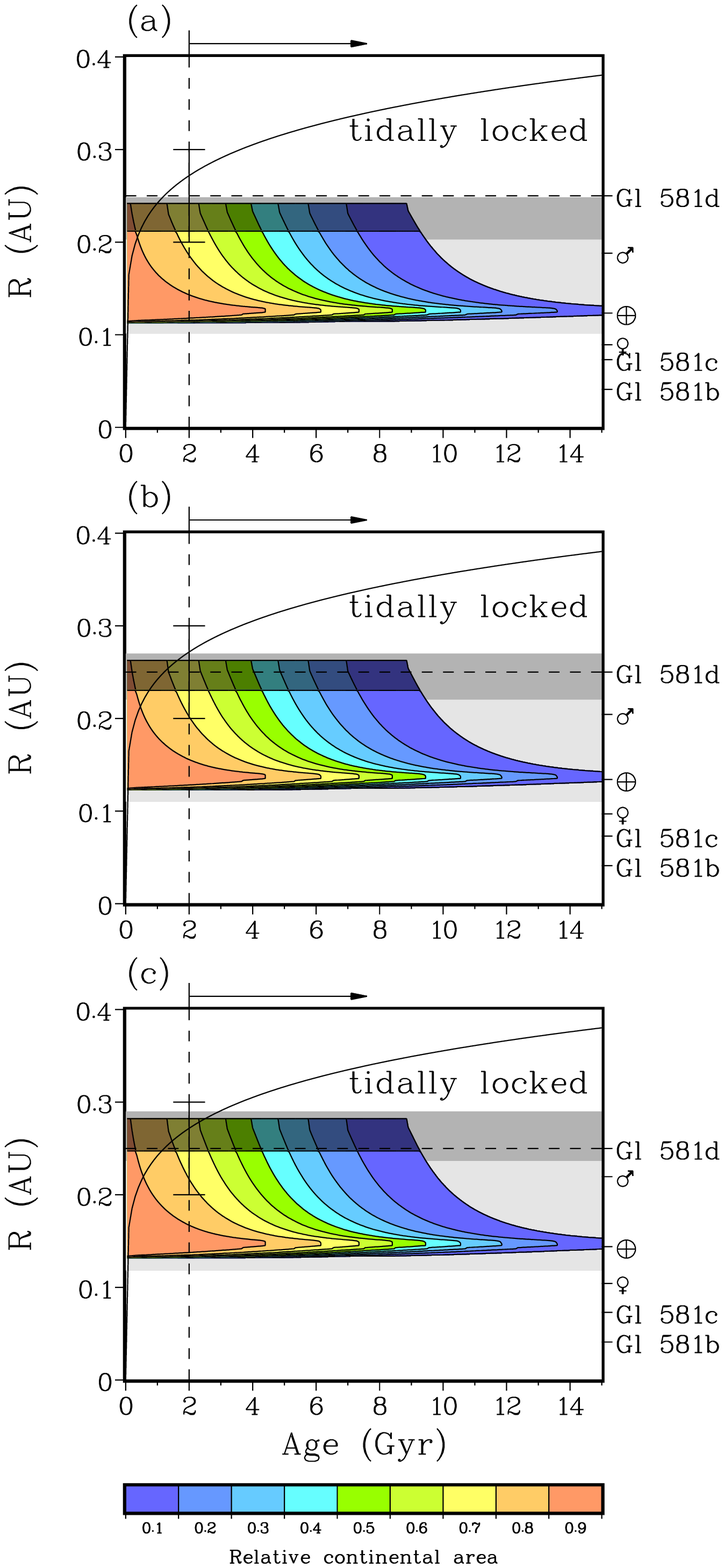}}
\caption{The pHZ of Gl 581 for a super-Earth ($M=8 M_\oplus$) with a relative
continental area varied from 0.1 to 0.9 and a fixed stellar luminosity of (a) $0.011L_\odot$,
(b) $0.013 L_\odot$, and (c) $0.015 L_\odot$ as a function of planetary age. The light colours
correspond to a maximum
CO$_2$ pressure of 5 bar, whereas the dark colours correspond to 10 bar.  For comparison,
the positions of Venus, Earth and Mars are shown scaled to the luminosity of Gl~581.
The light grey shaded area denotes the HZ calculated from Eq.~(\ref{hz_jones}), while the
dark shaded area corresponds to an extended outer limit following \cite{mischna00}.
The vertical bar at 2 Gyr denotes the range of distances due to the (possibly) eccentric
orbit.  The area below the solid black curve is affected by tidal locking.}
\label{fig3}
\end{figure}

The pHZ around Gl 581 for super-Earths with five and eight Earth masses has been calculated for
$L=0.011, 0.013$, and $0.015 L_\odot$.  The results for $5 M_\oplus$ are shown in Fig. \ref{fig2}a-c. 
The simulations have been carried out for a maximum CO$_2$ pressure of 5 bar (light colours) and 10 bar
(dark colours) neglecting the cooling effect of CO$_2$ clouds.  We assume that the maximum CO$_2$ pressure of the atmosphere is not limited by the total amount of carbon on the planet. Sufficient amounts of carbon are assumed to be always available for building up a CO$_2$ atmosphere of up to $P_{\mathrm{max}}=10$ bar. The biogenic enhancement factor of
weathering $\alpha_{\mathrm{bio}}$ has been set to one, i.e., a direct dependence of weathering on
biological productivity according to Eq.~(\ref{eq:enhance}) is neglected. Hence, the pHZ is calculated assuming a biotic enhancement of weathering similar to the modern biosphere on Earth. The tidal locking radius given by
Eq.~(\ref{tidal}) is also shown.  It is evident that both planets are well inside the tidal
locking radius, assuming a stellar age of at least 2 Gyr \citep{bonfils05}.
The inner boundary of the pHZ moves slightly outward,
whereas the outer boundary decreases nonlinearly with age.  Up to a critical age, the outer limit is constant
and is determined by the maximum CO$_2$ atmospheric pressure.  Beyond this age, the outer
boundary moves inward due to geodynamic effects.  At this point the source of carbon released into
the atmosphere is too low to prevent a freezing catastrophe. 
 \begin{table}
      \caption[]{Maximum outer limit $R_\mathrm{out}$ of the photosynthesis-sustaining habitable zone of Gl 581 for $L=0.013 L_\odot$
                 as a function of the maximum atmospheric CO$_2$ pressure $P_\mathrm{max}$ for three different
                 climate models.}
         \label{rout}
     $$
         \begin{array}{llll}
            \hline\hline
            \noalign{\smallskip}
             P_\mathrm{max} (\mathrm{bar}) &  R_\mathrm{out}^\mathrm{a} \mathrm{(AU)}
                                           &  R_\mathrm{out}^\mathrm{b} \mathrm{(AU)}
                                           &  R_\mathrm{out}^\mathrm{c} \mathrm{(AU)} \\
            \noalign{\smallskip}
            \hline
            \noalign{\smallskip}
            4   & 0.207 & 0.253 & 0.220 \\
            5   & 0.210 & 0.257 & 0.228 \\    
            6   & 0.213 & 0.260 & 0.236 \\
            7   & 0.216 & 0.263 & 0.242 \\
            8   & 0.219 & 0.265 & 0.247 \\
            9   & 0.221 & 0.267 & 0.252 \\
            10  & 0.223 & 0.269 & 0.261 \\
            \noalign{\smallskip}
            \hline
         \end{array}
     $$
\begin{list}{}{}
\item[$^{\mathrm{a}}$] \cite{chamberlain80}
\item[$^{\mathrm{b}}$] \cite{budyko82}
\item[$^{\mathrm{c}}$] \cite{williams98}
\end{list}
   \end{table}

The planet Gl 581c is clearly outside the habitable zone for all three luminosities.
The luminosity of the central star would have to be as low as $L=0.0045 L_\odot$
to yield habitable solutions for this planet.  It should be pointed out that Gl 581c is closer
to its parent star than Venus to the Sun, even with the stellar luminosity scaled accordingly.
The results for Gl 581d are much more encouraging (Fig.~\ref{fig3}a-c).  In particular, for a maximum CO$_2$
concentration of 10 bar, the planet is inside the pHZ for $L\ge 0.0117 L_\odot$, a value consistent
with the observed luminosity of Gl 581.  Assuming an Earth-like fraction of continental area $a_c=0.3$,
this planet would be habitable for a duration of 5.75 Gyr. For a water world ($a_c=0.1$) the duration is
extended to 9 Gyr, while for a  land world ($a_c=0.9$) the duration is shortened to 0.3 Gyr.

Due to the tidal locking and the planet's position near the outer edge of the pHz, the appearance of complex life is
rather unlikely on Gl~581d, i.e., $\Pi_{\mathrm{complex}}\equiv 0$.  Life is adversely affected by low temperatures,
small amounts of visible light, and insufficient shielding from intense flares.  Thus, the simulations have been
repeated with a parameterisation of weathering according to Eq.~(\ref{alpha}). The biogenic enhancement factor
$\alpha_{\mathrm{bio}}$ has been set to $3.6$ assuming that complex life amplifies weathering by this factor
\citep{vonbloh03b}.  We find that the duration of habitability for primitive life is extended to $8.8$ Gyr for
$a_c=0.3$.  In general, higher values of $\alpha_{\mathrm{bio}}$ extend the life span of the biosphere
\citep{lenton01,franck06}.  Incidentally, \cite{tarter07} argue that complex life might in principle be
possible on planets around M stars.

\cite{udry07} determined the orbital eccentricity of Gl 581c and Gl 581d as $0.16 \pm 0.07$
and $0.20 \pm 0.10$, respectively.  Assuming an eccentric orbit does not affect the result
that Gl 581c is outside the pHZ at all times; however, for Gl 581d, it would imply
that it leaves and re-enters the pHZ during its orbital motion for any of the assumed
model parameters (i.e., planetary climate model and stellar luminosity).  But this would
not thwart planetary habitability since a planet with a sufficiently dense atmosphere could
harbour life even if its orbit is temporarily outside the HZ, see \cite{williams02}.

However, it is noteworthy that the possible non-zero eccentricity values for the Gl 581
super-Earths are highly unlikely owing to the small number of radial-velocity
measurements by \cite{udry07}.  As a small number of measurements always tends to render
an overestimate in the deduced eccentricity, as well as a highly uncertain
error bar \citep{butler06}, a circular orbit is strongly preferred in this case,
as also done by \cite{udry07}.  In this case, the planetary orbit of Gl 581d is found to
stay inside of the pHZ all the time for most of the considered planetary climate models
and stellar luminosities.

The ultimate life span of a super-Earth is determined by the merging of the inner and outer
pHZ boundaries and depends on the planetary mass.  For a planet older than this ultimate life span,
no habitability is found.  An Earth-like
planet with $1 M_\oplus$ and a relative continental area of $0.3$ has an ultimate life span of 
$8.8$ Gyr, while super-Earth planets with $5 M_\oplus$ and $8 M_\oplus$ have ultimate life spans of
$11.1$ Gyr and $11.9$ Gyr, respectively.  The critical age and 
ultimate life span is found to decrease with the relative continental area
It is obvious that an almost completely ocean-covered planet (``water world")
has the highest likelihood of being habitable; see also previous models for 47~UMa by
\cite{franck03}.  However, for an age of 2 Gyr, habitability for Gl 581d is not constraint
by the outer edge of the pHZ for continental to total planetary surface ratios of less than 0.7.
In the case of a biogenic enhancement factor $\alpha_{\mathrm{bio}}=3.6$ for complex life,
habitability is maintained even for a land world with a continental to total planetary surface ratio
of $0.9$.

The simulations have been repeated
for different maximum CO$_2$ pressures $P_\mathrm{max}$ and three climate models (Tab.~\ref{rout}).
For the climate model by \cite{williams98}, Gl 581d is habitable for $P_\mathrm{max}\geq 9$ bar, for a
grey atmosphere model \citep{chamberlain80} no habitability can be found, while for the Budyko model
\citep{budyko82} (climate sensitivity 4 K/2 $\times$ CO$_2$) habitability is attained for
$P_\mathrm{max}\ge 4$ bar.

A planet with eight Earth masses
has more volatiles than an Earth size planet to build up such a dense atmosphere.  This prevents
the atmosphere from freezing out due to tidal locking.  In case of an eccentric orbit of Gl 581d
($e=0.2$), the planet is habitable for the entire luminosity range considered in this study, even if
the maximum CO$_2$ pressure is assumed as low as 5 bar.  
In conclusion, one might expect that life may have originated on Gl~581d.  The appearance of complex life,
however, is unlikely due to the rather adverse environmental conditions.  To get an ultimate answer to
the profound question of life on Gl 581d, we have to await future space missions such as the TPF/Darwin.
They will allow for the first time to attempt the detection of biomarkers \citep{grenfell07} in the
atmospheres of the two super-Earths around Gl 581.

\end{document}